\documentclass[conference,10pt]{IEEEtran}
\IEEEoverridecommandlockouts
\usepackage{cite}
\usepackage{amsmath,amssymb,amsfonts,mathrsfs}
\usepackage{algorithmic}
\usepackage{graphicx}
\usepackage{textcomp}
\usepackage{braket}
\def\BibTeX{{\rm B\kern-.05em{\sc i\kern-.025em b}\kern-.08em
    T\kern-.1667em\lower.7ex\hbox{E}\kern-.125emX}}
    
\UseRawInputEncoding

\begin{document}

\title{Entropy of the Canonical Occupancy (Macro) State in the
Quantum Measurement Theory
\\
}

\author{Arnaldo Spalvieri \\
Dipartimento di Elettronica, Informazione e Bioingegneria \\
Politecnico di Milano\\
}

\maketitle

\begin{abstract}

The paper analyzes the probability distribution of the occupancy
numbers and the entropy of a system at the equilibrium composed by
an arbitrary number of non-interacting bosons. The probability
distribution is derived both by tracing out the environment from a
bosonic eigenstate of the union of environment and system of
interest (the empirical approach) and by tracing out the
environment from the mixed state of the union of environment and
system of interest (the Bayesian approach). In the thermodynamic
limit, the two coincide and are equal to the multinomial
distribution. Furthermore, the paper proposes to identify the
physical entropy of the bosonic system with the Shannon entropy of
the occupancy numbers, fixing certain contradictions that arise in
the classical analysis of thermodynamic entropy. Finally, by
leveraging an information-theoretic inequality between the entropy
of the multinomial distribution and the entropy of the
multivariate hypergeometric distribution, Bayesianism  and
empiricism are integrated into a common ''infomechanical''
framework.

\vspace{0.5cm}

{\em Keywords:}\hspace{0.2cm}{\bf Occupancy Numbers, Multivariate
Hypergeometric Distribution, Multinomial Distribution,  Canonical
Typicality, Gibbs Correction Factor, Sackur-Tetrode Entropy
Formula.}

\end{abstract}

\section{Introduction}

The concept of physical entropy is controversial since the times
of Boltzmann and Gibbs.  The following passage of an interview
with Shannon, whose fundamental contribution to the understanding
of entropy in the information-theoretic sense is today widely
recognized also in physics \cite{bit}, can be found in
\cite{tribus}:

{\em My greatest concern was what to call it. I thought of calling
it 'information,' but the word was overly used, so I decided to
call it 'uncertainty.' When I discussed it with John von Neumann,
he had a better idea. Von Neumann told me, ''You should call it
entropy, for two reasons. In the first place your uncertainty
function has been used in statistical mechanics under that name,
so it already has a name. In the second place, and more important,
no one really knows what entropy really is, so in a debate you
will always have the advantage.''} \newline During the years, the
number of different interpretations and definitions of entropy has
grown, a recent collection of heterogeneous ''entropies'' being
reported in \cite{obsafra}.

In this fragmented and ambiguous context, one of the most debated
points is the relationship between information and physical
entropy. The controversial about the role of information in
physics plunge its roots in the famous thought experiments of
Maxwell and Szilard, and is still today object of discussion. From
the one side, Landauer claims in \cite{land2} that

{\em Information is a physical entity}. \newline From the other
side, Maroney writes in his thesis \cite{mt} that

{\em The Szilard Engine is unsuccessful as a paradigm of the
information-entropy link,} \newline and Maroney and Timpson
reiterate the same concept in \cite{maroneyrecent}:

{\em rejecting the claims that information is physical provides a
better basis for understanding the fertile relationship between
information theory and physics.}

Independently of the different opinions and arguments, it is a
matter of fact that entropy is the key quantity that allows to
derive all the thermodynamic macro-properties of many-particle
systems from the statistical properties of the constituent
microscopic particles.

One standard approach to the entropy in the classical
(non-quantum) regime is based on the Gibbs entropy of microstates
which, for systems at the equilibrium with the environment, is the
Shannon entropy of the probability distribution resulting from
entropy maximization under the constraints imposed on the system
by the environment that surrounds it. The principle of
maximization of entropy of microstates, often referred to as
MaxEnt, dates back to the seminal paper of Jaynes \cite{maxent}
and is inspired to Shannon's information-theoretic view of
entropy. Regrettably, the probability distribution of microstates
does not account for indistinguishability of particles. At the
same time, indistinguishability of particles is essential,
therefore it is imperative to incorporate it into the framework.
This is typically done by subtracting from the entropy of
microstates the Gibbsian correction term $\log(N!)$, where $N$ is
the number of system's particles.

The Gibbsian correction term enters the scene almost like a sudden
intervention, akin to a {\em deus ex machina}, but the subtraction
of $\log(N!)$ generates a glaring contradiction: when the entropy
of microstates is smaller than $\log(N!)$, the difference between
entropy of microstates and $\log(N!)$ becomes negative, as it
actually happens to the Sackur-Tetrode entropy formula at low
temperature-to-density ratio. Since the entropy of any random
variable is guaranteed to be always non-negative, entropy of
microstates minus $\log(N!)$ cannot be entropy. This pathology is
not surprising because, after the subtraction of $\log(N!)$, it is
no more specified the random variable whose ''entropy'' is
calculated, so ''entropy'' becomes a mere formula that is
attributed to an unspecified random variable that represents an
unspecified physical system. In this controversial situation,
after more than 100 years from its introduction, the Gibbsian
$\log(N!)$ is still today object of research and debate, see e.g.
\cite{sasa,tasaki}.


This paper recovers coherency of the picture by considering, in
place of the classical setting, the more general quantum setting,
where entropy is the von Neumann entropy of the mixed state of the
system. The mixed state of a system is commonly represented by a
density operator living in the Hilbert space ${\cal H}^{\otimes
N}$ obtained as the tensor product $N$ times  of the
single-particle Hilbert space ${\cal H}$. Exactly as it happens
with the entropy of microstates, the von Neumann entropy of this
density operator is appropriate for systems of distinguishable
particles, but it does not capture indistinguishability of
particles. We observe that, to capture indistinguishability of
particles of a bosonic system, i.e. a system whose state remains
unchanged when particles' indexes are permuted, an eigenbasis that
conveniently represents its quantum state is that of the vectors
of the occupancy numbers, giving rise to the so-called second
quantization formalism. The consideration of indistinguishability
of particles and, with it, of the occupancy numbers, leads us to
define the {\em bosonic} density operator i.e. the density
operator in system's bosonic subspace. Due to the inherent
capability of the bosonic density operator to capture
indistinguishability of particles, its von Neuman entropy {\em is}
the entropy of the system. This allows overcoming the standard
approach to entropy based on distinguishable particles and
resolves the puzzle of the Gibbsian $\log(N!)$, which, as shown in
this paper, now enters the scene through the main entrance
alongside other quantum correction terms ensuring non-negativity
of entropy.

Our exploration of the quantum apprach is situated within the
framework of contemporary quantum thermodynamics, whose origins
can be attributed to \cite{typpopescu,typgold}. The two cited
papers, which have deeply influenced the successive literature,
e.g. \cite{typbart,trends,itthermoreview,parrondo,deffner}, derive
the mixed state of a system at the equilibrium with the
environment by tracing out the environment from the universe,
that, in the thermal case, is the union of system and heat bath
that thermalizes the system. The analysis of \cite{typpopescu} and
\cite{typgold} shows that, as the number of particles of the
universe tends to infinity, the mixed state of the system is the
same, at least in the weak sense, both if the universe is in a
pure state or in a mixed state. Purity of the state of the
universe makes unnecessary the introduction of a statistical
ensemble of ''universes,'' overcoming the subjectivism that is
inherent in the Bayesian approach.


The outline of the paper is as follows. Section II introduces the
bosonic Hilbert subspace and its eigenbasis. In Section III, we
derive the bosonic density operator for the system under
consideration by tracing out the environment from the universe,
assuming that the universe is in a bosonic eigenstate. With this
assumption, we show that the probability distribution that weights
the projectors of the bosonic Hilbert subspace of the system is
the multivariate hypergeometric distribution. As the number of
bosons of the universe tends to infinity, the multivariate
hypergeometric distribution converges to the multinomial
distribution, which we identify as the canonical distribution of
the occupancy numbers. Section IV shows that, if, as in the
Bayesian approach, the universe is assumed to be in a mixed
bosonic state, then the distribution of the occupancy numbers of
the system is multinomial provided that the occupancy numbers of
the universe are multinomially distributed too. Section V
discusses the application of the mentioned probability
distributions to the entropy of physical systems and places our
work within the framework of quantum information theory, unveiling
an engaging connection between Bayesianism and empiricism in
physics. In section VI we discuss the results of the two previous
sections placing them within the framework of quantum measurement
theory. To illustrate the intrinsic capacity of the canonical
bosonic density operator to capture indistinguishability of
particles in systems at the equilibrium, Section VII shows that
its von Neumann entropy fits the entropy of the ideal gas in a
container. Section VIII sketches future application of our
approach to the Szilard engine. Finally, in Section IX we draw the
conclusions.

\section{The bosonic Hilbert subspace}

Let $\{{c} \in \mathbb{C} \}$, $\mathbb{C}=\{1,2, \cdots,
|\mathbb{C}|\}$, be the set of quantum numbers (the colors)
allowed to a boson, let $\{\bar{c}=(c_1,c_2, \cdots,c_N)\}$ (the
overline denotes vectors and their size is not explicitly
mentioned for brevity when it can be inferred from the context) be
the set of {\em microstates} of a system made by $N$
non-interacting bosons (the balls), and let $\{\ket{\bar{c}}\}$ be
the complete set of eigenstates that span the Hilbert space ${\cal
H}^{\otimes N}$ of the system. Let $\delta(\cdot)$ be the
indicator function and let
\[n_c(c_1,c_2, \cdots, c_N)\stackrel{\text{def}}{=} \sum_{i=1}^N \delta(c-c_i), \ c=1,2, \cdots,
|\mathbb{C}|,
\] be the number of bosons whose quantum number is ${c}$. In the following,
the dependency on $(c_1,c_2, \cdots c_N)$ of $n_c(c_1,c_2, \cdots
c_N)$ will be omitted when possible and  the vector of the
occupancy numbers ${\bar{n}}=(n_1,n_2, \cdots,n_{|\mathbb{C}|})$
will be called {\em occupancy macrostate}. The quantum state of
the system of indistinguishable bosons is
\begin{align}\ket{\bar{n}}=
\sum_{\bar{c} \in
\mathbb{C}^N(\bar{n})}\frac{\ket{\bar{c}}}{\sqrt{W(\bar{n})}},
\label{sdf}
\end{align}
see e.g. \cite{parisforthback} and chapter 7 of  \cite{kardar}.
In the above equation, $\mathbb{C}^N(\bar{n})$ is the set of the
$W(\bar{{n}})$ equiprobable microstates whose occupancy macrostate
is $\bar{n}$ and $W(\bar{n})$ is the multinomial coefficient,
\begin{align} W(\bar{{n}})
=\left\{
\begin{array}{cc}  \frac{N!}{\prod_{c=1}^{| \mathbb{C}|}n_c!}, &
n_c \geq 0, \ \forall \ c \in {\mathbb C}, \\
0,  &  \mbox{elsewhere},
\end{array} \right.
\nonumber \end{align} i.e. the number of distinct permutations of
the entries of anyone of the elements of $\mathbb{C}^N(\bar{n})$.

The set $\{\ket{\bar{n}}, {\bar{n}} \in \mathbb{N}\}$, with
\begin{align}  |\mathbb{N}|=\left(
\begin{array}{c} N+|\mathbb{C}|-1 \\
|\mathbb{C}|-1
\end{array} \right)
\nonumber \end{align} see \cite{cover}, is a complete set of
bosonic eigenstates for the bosonic Hilbert subspace of the
Hilbert space of the system.
\section{Empirical approach}

 We want to obtain the system's bosonic density operator $\hat{\nu}_{\bar{u}}$,
where $\bar{u}$ is the occupancy macrostate of the universe,
tracing out the environment made by $U-N$ bosons from the
projector $\ket{\bar{u}} \bra{\bar{u}}$ of the universe made by
$U$ bosons. The first step is to write $\ket{\bar{u}}$ in the form
of bosonic purification:
\begin{align}
\ket{\bar{u}}&= \sum_{\bar{n} \in {\mathbb
N},}\sum_{\bar{c}_{1}^{N} \in \mathbb{C}^{N}(\bar{n}),}
\sum_{\bar{c}_{N+1}^{U} \in
\mathbb{C}^{U-N}(\bar{u}-\bar{n})}\frac{\ket{\bar{c}_{1}^{N},
\bar{c}_{N+1}^{U}}}
{\sqrt{W(\bar{u})}}  \nonumber \\
&= \sum_{\bar{n} \in {\mathbb N}}
\sqrt{\frac{W(\bar{n})W(\bar{u}-\bar{n})}{W(\bar{u})}}
\ket{\bar{n},\bar{u}-\bar{n}}, \nonumber
\end{align}
where $\bar{c}_{i}^{j}$ means windowing of vector $\bar{c}$
between $i$ and $j$ and it is understood that the set
$\mathbb{C}^{N}(\bar{n})$ ($\mathbb{C}^{U-N}(\bar{u}-\bar{n})$) is
empty when one or more entries of $\bar{n}$ ($\bar{u}-\bar{n}$)
are negative. Observing that
\[\braket{\bar{u}-\bar{n}|\bar{u}-\bar{n}'}
\ket{\bar{n}} \bra{\bar{n}'}=\left\{
\begin{array}{cc} \hspace{-0.1cm} \ket{\bar{n}}
\bra{\bar{n}'}, & \hspace{-0.2cm} \bar{u}-\bar{n}=\bar{u}-\bar{n}',
\\
\hspace{-0.3cm} 0, & \hspace{-0.2cm} \bar{u}-\bar{n}
\neq \bar{u}-\bar{n}',\\
\end{array}\right.\]
and that the condition $\bar{u}-\bar{n}=\bar{u}-\bar{n}'$ forces
$\bar{n}=\bar{n}'$, we conclude that
\begin{align}
\hat{\nu}_{\bar{u}}&=\mbox{Tr}_{\hspace{0.05cm} {{\cal H}^{\otimes
(U-N)}}}\ket{\bar{u}} \bra{\bar{u}} \nonumber
\\ &
=\sum_{\bar{n} \in {\mathbb N}}\frac{W(\bar{u}-\bar{n})W(\bar{n})}
{W(\bar{u})}\ket{\bar{n}}\bra{\bar{n}}, \label{nu1}\end{align}
where $\mbox{Tr}_{\hspace{0.05cm} {{\cal H}^{\otimes (U-N)}}}$ is
the operator that traces out the Hilbert space of the environment.
If the set $\mathbb{C}^{N}(\bar{n})$
($\mathbb{C}^{U-N}(\bar{u}-\bar{n})$) is empty, then one or more
entries of $\bar{n}$ ($\bar{u}-\bar{n}$) are negative, in which
case the multinomial coefficient $W(\bar{n})$
($W(\bar{u}-\bar{n})$) is zero by definition. The fraction
appearing in (\ref{nu1}) is the multivariate hypergeometric
distribution, which is the distribution of the occupancy numbers
of colors in drawing without replacement $N$ balls out of an urn
containing $U$ balls with color occupancy numbers $\bar{u}$. In
many textbooks and papers, the multivariate hypergeometric
distribution is expressed by binomial coefficients as
\begin{align}
P_{\bar{u}}(\bar{n})=\frac{W(\bar{u}-\bar{n})W(\bar{n})}
{W(\bar{u})}=\frac{\prod_{c=1}^{| \mathbb{C}|}\left(
\begin{array}{c} u_c \\
n_c\\
\end{array} \right)}{\left(
\begin{array}{c} U \\
N\\
\end{array} \right)},
\label{mh}\end{align} where $P(\cdot)$ is the probability of the
random variable inside the round brackets and $_{\bar{u}}$ in the
subscript means that the probability distribution
$\{P_{\bar{u}}(\bar{n})\}$ depends on the vector of known
parameters ${\bar{u}}$.

In the thermodynamic limit the multivariate hypergeometric
distribution converges to the multinomial distribution, i.e. the
distribution of the occupancy numbers of colors in drawing with
replacement $N$ times a ball out of an urn containing colored
balls with relative frequency of color $c$ equal to $P(c)$:
\begin{align}\lim_{U \rightarrow
\infty}P_{\bar{u}}(\bar{n})&=W(\bar{n})\lim_{U \rightarrow
\infty}\frac{W(\bar{u}-\bar{n})}{W(\bar{u})}\nonumber \\ &=
W(\bar{n})\lim_{U \rightarrow \infty} \prod_{c=1}^{|
\mathbb{C}|}(U^{-1}u_c)^{n_c}\label{stirling}
\\ & =W(\bar{n})\prod_{c=1}^{| \mathbb{C}|} (P(c))^{n_c},\label{lln}
\end{align}
\begin{align}\lim_{U \rightarrow
\infty}\hat{\nu}_{\bar{u}}=\hat{\nu}=\sum_{\bar{n} \in {\mathbb
N}} W(\bar{n})\prod_{c=1}^{|\mathbb{C}|}
(P(c))^{n_c}\ket{\bar{n}}\bra{\bar{n}} ,\nonumber\end{align} where
(\ref{stirling}) is Stirling's formula, while, by regarding
${\bar{u}}$ as the result of a PVM operated on the universe,
equation (\ref{lln}) is recognized to be the Law of Large Numbers
(LLN) for the empirical one-particle distribution.  {\em
Concentration inequalities} that bound the probability of
deviations of the empirical probability from its expectation, that
is the probability of occurrence of {\em non-typical} bosonic
eigenstates of the universe, can be found in \cite{greene} for
hypergeometrically distributed random variables, in \cite{agrawal}
for multinomially distributed random vectors. Paper \cite{lottery}
demonstrates that the multinomial distribution is the maximum
entropy distribution of the occupancy numbers with constrained
one-particle distribution. See also \cite{pnas,zupa} for the
multinomial distribution in statistical mechanics.

By equiprobability of the disjoint microstates belonging to the
same occupancy macrostate, we see that, in (\ref{mh}),
\begin{align}\frac{W(\bar{u}-\bar{n}(c_1,c_2, \cdots, c_N))}
{W(\bar{u})}=P_{\bar{u}}(c_1,c_2, \cdots, c_N).\nonumber
\end{align} Clearly
\begin{align}\lim_{U \rightarrow \infty}P_{\bar{u}}(c_1,c_2, \cdots,
c_N)&=\prod_{c=1}^{|\mathbb{C}|} (P(c))^{n_c(c_1,c_2, \cdots,
c_N)}\nonumber \\ &=\prod_{i=1}^{N} P(c_i),\nonumber \end{align}
which explicitly shows that, in the thermodynamic limit,
microstates outcoming from the mentioned PVM  becomes independent
and identically distributed random variables. What happens is that
the LLN cancels the dependencies inside the joint distribution of
microstates induced by the constraints imposed on the colors of
the $N$ balls by the occupancy numbers $\bar{u}$ of the universe
in drawing without replacement. Furthermore, the probability
distribution $\{\prod_{i=1}^{N} P(c_{i})\}$ also becomes
independent of the specific result $\bar{u}$ of the PVM operated
on the universe, in the sense that the empirical one-particle
distribution $\{U^{-1}u_c\}$ for $U \rightarrow \infty$ is the
same {\em for almost every bosonic eigenstate of the universe},
i.e., for the {\em typical} bosonic eigenstates of the universe.

\begin{figure}[!h]
\begin{center}
\includegraphics[ scale=0.7]{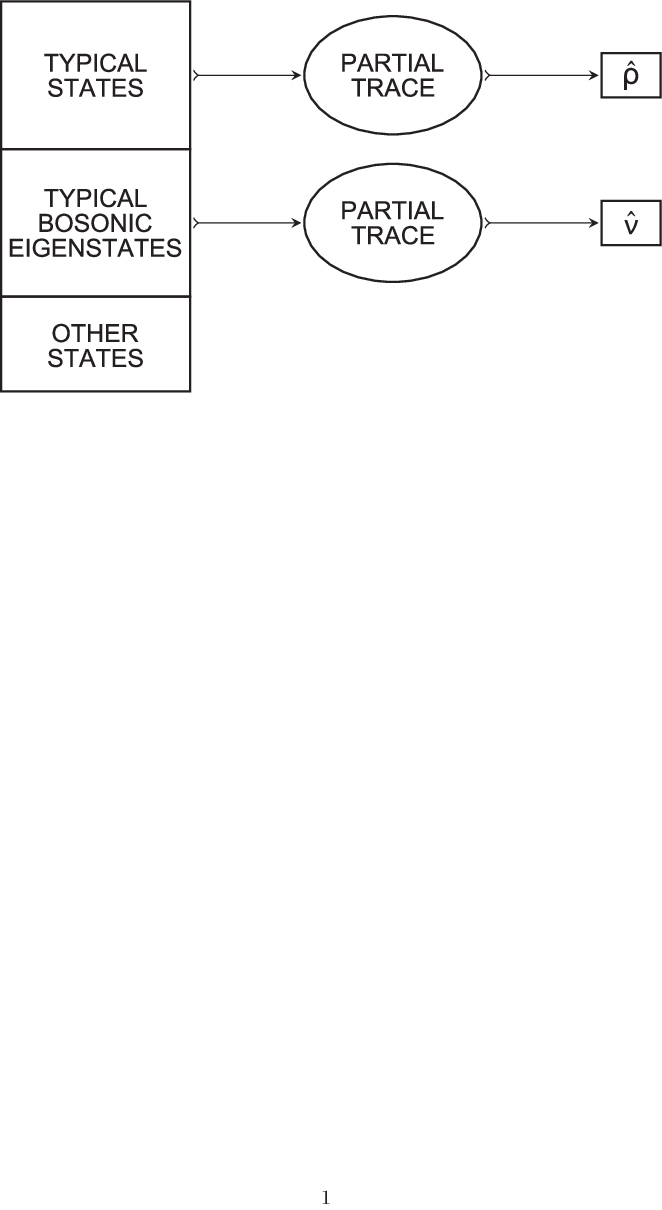}
\end{center}
\vspace*{-9.5cm} \caption{The box on the left represents the pure
states of the universe. Although in the thermodynamic limit almost
every pure state of the universe is typical, still there is space
for infinitely many typical bosonic eigenstates and for infinitely
many other pure states that are neither typical states nor typical
bosonic eigenstates, among which are all the eigenstates of ${\cal
H}^{\otimes U}$.} \label{fig}
\end{figure}

Independency and identical distribution of the eigenstates of the
individual bosons of the system lead to the following density
operator in ${\cal H}^{\otimes N}$
\begin{align}\hat{\rho}&=\sum_{\bar{c} \in
\mathbb{C}^N}\prod_{i=1}^{N} P(c_i)\ket{\bar{c}}\bra{\bar{c}}
,\nonumber
\end{align}
that we claim to be the canonical density operator in ${\cal
H}^{\otimes N}$. Papers \cite{typpopescu} and \cite{typgold} claim
that system's density operator in ${\cal H}^{\otimes N}$ converges
to the canonical density operator as $U \rightarrow \infty$  {\em
for almost every pure state of the universe,} i.e. for the {\em
typical} pure states of the universe. For this reason convergence
to the canonical state is called {\em canonical typicality} in
\cite{typgold}. Typicality, which can be intended in various
senses, is often invoked in statistical mechanics. For instance,
paper \cite{spalvieri} uses the properties of the
information-theoretic typical set to characterize weak convergence
of microstates to equiprobability in the context of classical
statistical mechanics. We point out that the claim of
\cite{typpopescu} and \cite{typgold} is compatible with our claim
of convergence to the bosonic canonical state if we do not pretend
that typical bosonic eigenstates are typical states, as
illustrated in Fig. \ref{fig}.

\section{Bayesian approach}

In the Bayesian approach, the vector of known parameters $\bar{u}$
becomes a random vector whose probability distribution, the
Bayesian {\em prior}, characterizes the bosonic density operator
of the universe. In the Bayesian setting, the hypergeometric
multivariate distribution of the density operator
$\hat{\nu}_{\bar{u}}$ of (\ref{nu1}) is the Bayesian {\em
likelihood}, i.e. a conditional distribution where the random
condition is the random vector of parameters $\bar{u}$. The
probability distribution of the density operator of the system,
the Bayesian {\em marginal}, is obtained by tracing out the
environment from the bosonic density operator of the universe. The
Bayesian approach is self-consistent when the prior is
multinomial, because the marginal turns out to be multinomial too,
whichever is the number of particles of the universe:
\begin{align}\hat{\nu}&=\sum_{\bar{u} \in {\mathbb U}}
W(\bar{u})\prod_{c=1}^{|\mathbb{C}|} (P(c))^{u_c}
\mbox{Tr}_{\hspace{0.05cm} {{\cal H}^{\otimes
(U-N)}}}\ket{\bar{u}} \bra{\bar{u}} \nonumber
\\ &= \sum_{\bar{u} \in {\mathbb U}}
W(\bar{u})\prod_{c=1}^{|\mathbb{C}|} (P(c))^{u_c}
\hat{\nu}_{\bar{u}} \nonumber \\ &=\sum_{\bar{n} \in {\mathbb N}}
W(\bar{n})\prod_{c=1}^{|\mathbb{C}|}
(P(c))^{n_c}\ket{\bar{n}}\bra{\bar{n}}, \label{bayes}\end{align}
where the third equality is obtained by substituting (\ref{nu1})
in the second line and manipulating. The comparison between the
Bayesian (\ref{bayes}) and the empirical (\ref{nu1}) clarifies the
relationship between the two approaches. The Bayesian approach
captures the pre-measurement randomness of the mixed state of the
universe by a mathematical procedure based on constrained entropy
maximization. Specifically, when the constraint is the
temperature, the result of Jaynes' MaxEnt approach \cite{maxent}
is that $\{\prod_{i=1}^{N} P(c_{i})\}$ is the Boltzmann
distribution. Instead, the empirical approach renounces to
mathematics and imports the result $\bar{u}$ of the measurement of
the occupancy numbers of the universe in the multivariate
hypergeometric distribution of the occupancy numbers of a system
that is separated from the universe after the measurement operated
on the universe.

\section{Quantum entropy and quantum information}

Since $\hat{\nu}_{\bar{u}}$ ($\hat{\nu}$) is an
eigendecomposition, its von Neumann entropy is equal to the
Shannon entropy of the multivariate hypergeometric (multinomial)
distribution. For this reason, in the following we make no
distinction between the von Neumann entropy and the Shannon
entropy, calling it simply entropy. It is worth emphasizing that
here we completely skip the notion of phase space, leading to the
{\em exact} probability distribution of the quantum occupancy
numbers, and, as a consequence, to the exact entropy. Conversely,
the standard phase space approach inherently leads to
approximations to entropy, that ask for improvement at low
temperature/density ratio, see e.g. \cite{psapprox}, still
remaining approximations.

The entropy of the multivariate hypergeometric distribution,
which, owing to the distribution's symmetry, is equal to the
entropy of both the system and the environment after their
separation, is
\begin{align}
&S({\hat{\nu}_u})  = \log(W(\bar{u}))-E\{\log(W(\bar{u}-\bar{n})
\}-E\{\log(W(\bar{n}))\}\nonumber
\\ &=-E\{\log(P_{\bar
u}(\bar{c}))\}-\log(N!) + \sum_{c=1}^{|{\mathbb C}|}
E\{{\log(n_{c}!)}\}\label{difficult1},
\end{align}
where $S(\cdot)$ is the von Neuman entropy of the density operator
inside the round brackets, $E\{\cdot\}$ is the classical
expectation computed over the probability distribution of the
random variable inside the expectation, the base of the logarithm
is Euler's number, and the Boltzmann constant in front of the
logarithm is omitted for brevity. The term $-E\{\log(P_{\bar
u}(\bar{c}))\}$ is the Gibbs entropy of microstates, i.e. the
entropy of the system of distinguishable particles, while the term
$-E\{\log(W(\bar{n}))\}$ is due to indistinguishability of
particles which, in the average, prevents the access to
$E\{\log(W(\bar{n}))\}$ units of information. Since
$(W(\bar{n}))^{-1}$ is the conditional probability of microstates
given the macrostate, its expectation is the conditional Shannon
entropy of microstates given the macrostate, so $S({\hat{\nu}_u})$
is the mutual information between microstates and macrostates. The
term $\log(N!)$ in (\ref{difficult1}) was introduced by Gibbs to
force compatibility between the non-quantized phase-space
(differential) entropy of microstates and the physical entropy of
systems of indistinguishable particles. We observe that, while the
probability that two or more classical particles have the same
position and momentum in the phase space is zero because position
and momentum are dense variables, the probability that two or more
quantum particles occupy the same quantum state is not zero. This
non-zero probability is captured by the sum of expectations in
(\ref{difficult1}). As the entropy of microstates becomes lower
and lower, this sum becomes closer and closer to $\log(N!)$, till
becoming equal to $\log(N!)$ when all the particles occupy the
ground state. This prevents system's entropy to become negative
also when the entropy of microstates becomes vanishingly small.

In the canonical case, the entropy of the bosonic density operator
is the entropy of the multinomial distribution:
\begin{align}
&S({\hat{\nu}})  = -NE\{\log(P(c))\}-E\{\log(W(\bar{n}))\},
\label{entropynu}
\end{align}
see \cite{me} and \cite{mahdi} for the calculation of the above
entropy, see also \cite{zupa} for approximations to the entropy of
the multinomial distribution in the context of statistical
mechanics. Equation (\ref{entropynu}) is equation (11) of
\cite{zupa}, where the authors call the entropy of the
distribution of the occupancy numbers {\em entropy fluctuations}.
Apart of certain exceptions, the authors of \cite{zupa} consider
these ''entropy fluctuations'' negligible compared to the
''entropy'' of the system, failing to recognize that the entropy
of the occupancy numbers {\em is} the thermodynamic entropy of a
system of indistinguishable particles.

The following inequalities sandwich the Boltzmann entropy
$\log(W(E\{\bar{n}\}))$ between the two terms in the right hand
side of (\ref{entropynu}):
\begin{align}
-NE\{\log(P(c))\} &\geq \log(W(E\{\bar{n}\}))\label{stirling1} \\
&  \geq  E\{ \log(W(\bar{n}))\} ,\nonumber
\end{align}
where, with some abuse of notation, the factorials of the real
numbers in the denominator of $W(E\{\bar{n}\})$ are intended as
$x!=\Gamma(x+1),$ where $\Gamma(\cdot)$ is the Gamma function. The
first inequality is (11.22) of \cite{cover}, the second inequality
is obtained by applying the Jensen inequality
\[E\{ f(n_c) \} \geq f(E\{ n_c \}), \ \forall \ c \in {\mathbb C},\]
to the convex (upward) function $f(n_c)=\log(n_c!)$. In
statistical mechanics it is standard to derive from Stirling's
formula an approximation between the two terms of
(\ref{stirling1}).

Note that, if we pretend that entropy is a variable of state, then
the probability distribution of the occupancy numbers must depend
only on the state of the system. However, the multivariate
hypergeometric distribution depends also on the state of the
universe. The dependency becomes weaker and weaker as the number
of particles of the universe tends to infinity, but it remains
that this makes the empirical approach incompatible with the
notion of entropy as variable of state. In conclusion, entropy can
be a variable of state only if we accept the Bayesian approach.

We hereafter introduce the {\em empirical} information, sketching
a new engaging connection between the Bayesian and the empirical
approaches to information in quantum measurement theory. Let us
regard the PVM operated on the universe as a POVM operated on the
system. The difference between the entropy of the multinomial
Bayesian marginal and the expectation over the multinomial
Bayesian prior $\{P(\bar{u})\}$ of the entropy of the multivariate
hypergeometric Bayesian likelihood of the system is equal to the
Holevo upper bound $\chi$ above the accessible quantum information
that any POVM can achieve \cite{ikemike}:
\begin{align}&\chi=S({\hat{\nu}})-\sum_{\bar{u}\in {\mathbb U}}
P(\bar{u})S({\hat{\nu}_{\bar{u}}})\geq 0.\label{chi}\end{align}

Whichever is $U$, in place of the above Bayesian information one
could consistently consider the following empirical information,
that does not need the definition of a prior:
\begin{equation}
S(\hat{\nu}_{\rm{emp}})-S(\hat{\nu}_{\bar{u}}) \geq
0,\label{diff}\end{equation} where $S(\hat{\nu}_{\rm{emp}})$ is
the Shannon entropy of the empirical multinomial distribution
$\{P_{\rm{emp}}(\bar{n})\}$ of the occupancy numbers of the
system,
\[P_{\rm{emp}}(\bar{n})=W(\bar{n})\prod_{c=1}^{|\mathbb{C}|}(U^{-1}u_c)^{n_c}.\] In
(\ref{diff}), the one-particle probability distribution
$\{U^{-1}u_c\}$ is the same for the multivariate hypergeometric
distribution and for the empirical multinomial distribution,
therefore the inequality is guaranteed by the maximum entropy
property of the multinomial distribution demonstrated in
\cite{lottery}. The difference (\ref{diff}) is the ''empirical
information''  brought by PVM operated on the universe about the
system.  As $U \rightarrow \infty$, both $S(\hat{\nu}_{\bar{u}})$
and $S(\hat{\nu}_{\rm{emp}})$ tend to $S(\hat{\nu})$ and both the
Bayesian information and the empirical information tend to zero.
If, after the POVM, a PVM is operated on the system, the total
information brought by the two measurements is $S(\hat{\nu})$ in
the Bayesian approach, $S(\hat{\nu}_{\rm{emp}})$ in the empirical
approach. As $U \rightarrow \infty$, the total empirical
information becomes equal to the total Bayesian information.

\section{Collapse of entropy and production of information in the theory of quantum measurement}

In the endless debate that opposes Bayesians to empiricists, the
Bayesian position is synthesized by Jaynes in \cite{maxent} as
follows: {\em We frankly recognize that the probabilities involved
in prediction based on partial information can have only a
subjective significance, and that the situation cannot be altered
by the device of inventing a fictitious ensemble, even though this
enables us to give the probabilities a frequency
interpretation.}\newline  In Jaynes' vision, the probability
distribution, in our case the MaxEnt multinomial probability
distribution, is derived from a mathematical procedure in which
the physician {\em trusts} for some reason. For instance, in the
thermal case the trusted mathematical procedure leads to the
Boltzmann distribution of system's microstates.

We hereafter discuss another kind of subjectivity that both the
empirical and the Bayesian entropies directly inherit from the
theory of quantum measurement. Suppose that an experimenter, after
having separated the system from the environment, performs a PVM
on the system to gain information about the state of the system.
After the measurement, the pre-measurement mixed state of the
system collapses into the eigenstate where the PVM leaves the
system, the probability distribution that the experimenter
attributes to the system before the measurement, be it the
multinomial distribution or the multivariate hypergeometric
distribution, collapses to an indicator function, system's entropy
collapses into zero and the pre-measurement entropy becomes
information gained by the experimenter. According to the theory of
quantum measurement, all these changes impact only the probability
distribution that the experimenter attributes to the system, while
they do not impact neither the system nor the surrounding
environment, including a second experimenter that measures the
system after the first one. Actually, provided that between an
experiment and the successive system's particles do not interact
with the environment and between them, the second experimenter
will find the same occupancy numbers found by the first one. We
see therefore that, while system's state is objective and the
experiment is deterministically repeatable, entropy is inherently
subjective. Subjectivity comes directly from the theory of quantum
measurement: the pre-measurement uncertainty of the experimenter,
what we call {\em entropy}, is transformed by the collapse of the
mixed state of the system into one of its eigenstates into
post-measurement information available to the experimenter. This
transformation is subjective because it impacts only the
experimenter, as all the properties of the system perceived by the
rest of the world remain unchanged. Meanwhile, for the second
experimenter the system is still in the mixed state, so he is
still uncertain about the eigenstate that will be detected by his
measurement: for him, the system still has entropy. The situation
is analogous to that of information transmission, where a
regenerative repeater receives, detects, and then transmits a
message. In this entire process the message (system) does not
change, what changes is the state of the repeater (experimenter)
that from receiver (measurer) becomes transmitter (preparer).
The picture doesn't change so much if, in place of a PVM, the
experimenter performs a POVM. The only difference is that the
post-measurement distribution is no more an indicator function,
but also in this case the post-measurement entropy is not greater
than the pre-measurement entropy, the difference between the two
being equal to the information that the POVM brings to the
experimenter about the system.

Suppose that the experimenter wants to know how much information
he gained from the measurement, or, equivalently, how much
uncertainty (entropy) the experiment removed. In this case he must
attribute pre-measurement and post-measurement probability
distributions to the system. Consider the case where the full
information about the system is gained by optionally performing
one or more POVMs followed by a PVM that leaves the system in an
eigenstate. The post-measurement distribution is an indicator
function both in the Bayesian approach and in the empirical
approach. The difference between the two approaches is in the
pre-measurement distribution. In the Bayesian approach the
pre-measurement distribution is the maximum entropy distribution
and the full information about the system can be gained by
performing a PVM on the system, or, equivalently, by performing a
PVM on the universe which partially collapses system's entropy
from that of the multinomial distribution to that of the
multivariate hypergeometric distribution, followed by a PVM on the
system that collapses the entropy of the multivariate
hypergeometric distribution to zero. In the empirical approach,
the experimenter refuses to define a pre-measurement distribution,
so the PVM operated on the universe partially collapses system's
entropy from an undefined value to the entropy of the multivariate
hypergeometric distribution, and the successive PVM operated on
the system collapses entropy to zero. This situation is
unsatisfactory, as it is clearly seen by looking at the case where
the system is the universe. In this case, entropy collapses from
an undefined value to zero, and information raises from zero to an
undefined value. Although we all recognize that the universe has
entropy and that, through the measurement, the experimenter gains
information about it, due to his agnostic position about the
pre-measurement distribution, the empiricist cannot quantify
entropy (information). A way out from this uncomfortable situation
is offered to the empiricist by our proposed empirical
information. In place of defining a prior, the empiricist can
assume a multinomial model based on the empiric one-particle
distribution obtained from the PVM operated on the universe. When
regarded as a POVM operated on the system, this measurement
partially collapses system's entropy from the entropy of the
empirical multinomial distribution to the entropy of the
multivariate hypergeometric distribution. The successive PVM
collapses entropy to zero, so the total information gained by the
two measurements is the entropy of the empirical multinomial
distribution.

Before concluding this section we return on the {\em vexata
quaestio} of Bayesianism against empiricism. We have seen that,
while the Bayesian assumes a multinomial model based on an
one-particle distribution that he derives from a mathematical
model, the empiricist can assume a multinomial model based on an
one-particle distribution derived from the experiment as a way out
from the issue of quantifying the entropy of the universe. In both
cases, the maximum entropy property of the multinomial model saves
non-negativity of information and, with it, of entropy. Despite
this analogy, Bayesianism and empiricism remain inherently
different, as it can be seen by considering the thermal case and
an universe with few particles. In this case this author finds
more comfortable the position of the Bayesian. Actually, while the
Bayesian, who trusts in the Boltzmann distribution, has to infer
from the measurement made on the few particles of the universe
only the temperature, the empiricist, who trusts only in the
experimental evidence, has to infer from that measurement the
entire one-particle probability distribution. From this
perspective, we can say that the inference performed by the
Bayesian is parametric, while the inference made by the empiricist
is non-parametric. It is well known that, provided that the
parametric model fits the system under investigation, with small
sample size the inference of few parameters, in the thermal case,
only the temperature, will be of better quality than the inference
of many unknowns.


%
%
%
%
%
%
%

\section{Entropy of the ideal gas in a container}

In the case of an ideal monoatomic gas in a cubic container of
side $L$, one particle of the gas is modelled as a quantum
''particle in a box'' with three degrees of freedom, whose energy
eigenvalues with aperiodic boundary conditions are
\begin{equation}\epsilon_{c}=
(c_x^2+c^2_y+c^2_z) \frac{h^2}{8 m L^2},\nonumber
\end{equation}
where $c$ consists of the three quantum numbers $(c_x,c_y,c_z)$,
$m$ is the mass of the particle and $h=6.626 \cdot 10^{-34} \ \
\mbox{J} \cdot \mbox{s}$ is the Planck constant. The one-particle
Boltzmann distribution for a gas at the thermal equilibrium at
temperature $T$ Kelvin degrees is
\begin{equation}P(c)=Z^{-1}
e^{-\epsilon_{c}/k_BT},\nonumber \end{equation} and the associated
multinomial distribution of the occupancy numbers for a gas of $N$
particles is
\begin{equation}P(\bar{n})=W(\bar{n})Z^{-N}\prod_{c=1}^{|{\mathbb C}|}
e^{-n_c\epsilon_c/k_BT},\nonumber \end{equation} where $k_B = 1.38
\cdot 10^{-23}$ J/K is the Boltzmann constant and $Z$ is the
one-particle partition function:
\begin{equation}Z=\sum_{c=1}^{|{\mathbb C}|}e^{- \epsilon_c/k_BT}. \nonumber
\end{equation}

When the temperature-to-density ratio is high, it becomes possible
to employ two approximations. In the first one, the partition
function is approximated to an integral, see eqn. 19.54 of
\cite{sek}, leading to
\begin{align}
-NE\{\log(P(c))\}&\approx \frac{3N}{2} \left(1+\log \left(\frac{2
\pi m k_B T L^2}{h^2}\right)\right) \nonumber.
\end{align}
In the second one, with the idea that the probability that two or
more particles occupy the same state is negligible, the
denominator of the multinomial coefficient is ignored and, for
large number of particles, the logarithm of the numerator
$\log(N!)$ is approximated to $N\log(N)-N$, leading to
\begin{align}
-E\{\log(W(\bar{n}))\}&\approx -N\log(N)+N \nonumber.
\end{align}
Plugging the two approximations in (\ref{entropynu}) one gets
textbook Sackur-Tetrode formula:
\begin{align}
S(\hat{\nu})\approx N\left(\log \left( \frac{L^3}{N}\left(\frac{2
\pi m k_B T}{h^2 }\right)^{\frac{3}{2}}\right)+\frac{5}{2}
\right). \nonumber
\end{align}

Note that, as already mentioned in the introduction, the exact
entropy of the multinomial distribution (\ref{entropynu}) is
guaranteed to be non-negative, while the Sackur-Tetrode formula
becomes negative at low temperature-to-density ratio, where the
two mentioned approximations do not hold, see
\cite{spalvarx1,spalvarx2} for details.

\section{Future work}

We hereafter sketch the application of our proposed approach to
the state of Szilard engine after the insertion of the piston,
letting the complete analysis of the Szilard cycle to future work.
In the case of $N$ particles, after the insertion of the piston
the number $b$ of particles that are found in one of the two
sub-container of volume $V'$ is a binomial random variable
$(N,V'/V)$, where $V$ is the total volume. The probability
distribution of the occupancy numbers is
\begin{align}
P(\bar{n}',\bar{n}'')=\sum_{b=0}^N
P(\bar{n}'|b)P(\bar{n}''|b)P(b), \nonumber
\end{align} where $\{P(\bar{n}'|b)\}$
($\{P(\bar{n}''|b)\}$) is the probability distribution of the
occupancy numbers of a gas with $b$ ($N-b$) particles in the
sub-container of volume $V'$ ($V-V'$). For $N=1$ and $V'=V/2$, the
entropy after the insertion of the wall is
\begin{align}\log(2)-\sum_{c=1}^{|\mathbb{C}|}P(c)\log(P(c)),
\nonumber\end{align} where the famous $\log(2)$ of Landauer
\cite{land} comes from the binary equiprobable random variable $b$
and $\{P(c)\}$ is the probability distribution of one particle in
a box of volume $V'$. We have evaluated the partition function of
the Boltzmann distribution with the parametrization of
\cite{aydin}, that is mass of the particle $m=9.11 \cdot 10^{-31}$
kg, temperature $T=300 $ K, and one-dimensional box of size $L=20
\cdot 10^{-9}$ m. We obtain that the entropy of the single
particle with one degree of freedom before the insertion of the
piston is $1.988$ in $k_B$ units, while with size of the
one-dimensional box equal to $10 \cdot 10^{-9}$ m, that is, after
the insertion of the piston, the entropy in $k_B$ units is
$1.243$, leading to the difference $1.988-0.693-1.243=0.052$, in
excellent agreement with the entropy fall shown in Fig. 3 of
\cite{aydin}, where the result is derived by the phase space
approach.

\section{Conclusion}

Entropy is a macroscopic property of a physical system and, at the
same time, it is a mathematical property of randomness. As such,
it {\em must} be a property of the randomness of system's
macrostates. However, despite the necessity of introducing the
somewhat ad hoc term $-\log(N!)$ as a kind of {\em deus ex
machina}, from Boltzmann to the present day the ''entropy'' of a
system at the equilibrium is commonly intended as the entropy of
microstates, be it the Gibbs entropy or, to some extents, the
Boltzmann entropy $\log(W(E\{\bar{n}\})$. This author has come to
the conclusion that the misunderstanding arises from the following
two conceptual errors.

The first error consists in regarding $\log(W(\bar{n}))$ as the
''entropy of a macrostate.'' Even an exceptionally deep author as
Jaynes writes in \cite{jgibbs}: {\em To emphasize this, note that
a ''thermodynamic state'' denoted by $X=\{X_1, \cdots, X_n\}$
defines a large class $C(X)$ of microstates compatible with $X$.
Boltzmann, Planck, and Einstein showed that we may interpret the
entropy of a macrostate as $S(X) = k log(W(C))$, where $W(C)$ is
the phase volume occupied by all the microstates in the chosen
reference class $C$.} This misconception is pervasive in the
entire statistical mechanics. The influential authors of
\cite{goldgibbs} attribute to Einstein the idea that an individual
macrostate can have entropy: {\em In fact, already Einstein (1914,
Eq. (4a)) argued that the entropy of a macro state should be
proportional to the log of the ''number of elementary quantum
states" compatible with that macro state...} . But entropy is a
property of a statistical ensemble, therefore one individual
macrostate cannot have entropy. Moreover, if by entropy of the
macrostate we mean the entropy of the microstates that constitute
it, then we must acknowledge that this entropy has no physical
significance because the {\em elementary quantum state compatible
with that macro state} is physically meaningless when particles
are indistinguishable. This misconception is so widespread that
can be found also in standard textbooks. We hereafter quote a
passage from the introduction to chapter 16 of \cite{sek}: {\em
specification of a macrostate constitutes incomplete information}.
In absence of a formal definition of information, this statement
risks to become misleading. Actually, we have shown that the
entropy of the occupancy numbers  {\em is} the complete
information about the system because the entropy of microstates
belonging to the same macrostate, due to indistinguishability of
particles, is not informative, as it is subtracted in
(\ref{entropynu}) to the entropy of microstates.

The second error is the lack of consideration of the absolute
randomness of macrostates. Navigating through the literature of
statistical mechanics, you may come across passages like the one
of \cite{pnas} that is hereafter reported: {\em A crucial
observation in statistical mechanics is that the distribution of
all macrostate variables gets sharply peaked and narrow as system
size N increases. ... In the limit $N \rightarrow \infty$ the
probability of measuring a macrostate becomes a Dirac delta...}
Clearly, the quoted statement is wrong, because the {\em absolute}
width of the probability distribution of the occupancy numbers,
and, more generally, of any macroscopic observable as for instance
system's energy, becomes broader and broader as the number of
particles grows, while, by the LLN, its {\em relative} width, i.e.
the width compared to the mean value, sharpens. Certainly, it is
apparent that
\[\lim_{N \rightarrow \infty}N^{-1}S(\hat{\nu})=0,\]
which shows that the relative randomness, i.e. the randomness per
particle, becomes vanishingly small as $N \rightarrow \infty$.
Meanwhile, the absolute randomness, represented by $S(\hat{\nu})$,
increases as $N$ becomes larger and larger.

In the end, the lack of formal specification of the physical role
of microstates, together with the lack of consideration of the
absolute randomness of macrostates, lead to the questionable
belief that all the properties of a system of a large numbers of
particles, including ''entropy,'' depend on the macrostate
$E\{\bar{n}\}$, because it contains ''the overwhelming majority''
of microstates. From this standpoint, the subtraction of
$\log(N!)$ is seen as a technical maneuver that becomes necessary
to circumvent the challenge posed by the indistinguishability of
particles. Note that the idea itself that one macrostate can
contain the overwhelming majority of microstates is inherently
questionable, because the ratio
$|\mathbb{C}^N(\bar{n})|/|\mathbb{C}^N|$ tends to zero whichever
is $\bar{n}$ as $N \rightarrow \infty$.

In conclusion, this paper made clear that, since the quantum state
of a system of indistinguishable particles (bosons) is completely
specified by the occupancy numbers of the quantum states allowed
to system's particles, the entropy the physical system is the
Shannon entropy of the random occupancy numbers $\bar{n}$, which
is obtained by subtracting the expectation of $\log(W(\bar{n}))$
to the entropy of microstates. Recognizing that
$E\{\log(W(\bar{n}))\}$ is the conditional Shannon entropy of
microstates given the macrostate, we equivalently express the
above concept by saying that the entropy of the physical system is
equal to the mutual information between microstates and
macrostates.

\end{document}